\documentclass[aps,prd,twocolumn,showkeys,showpacs,nofootinbib]{revtex4}
\usepackage{latexsym}
\usepackage{amsmath,amsfonts}
\usepackage{amsbsy}
\usepackage{mathrsfs}
\usepackage{color}
\usepackage{psfrag}
\usepackage{enumerate}
\usepackage{amsmath,amssymb,calc,amsfonts}
\usepackage{latexsym}
\usepackage{graphicx,calc,epsfig}
\def\ut#1{\rlap{\lower1ex\hbox{$\sim$}}#1{}}

\newcommand{\be}{\nopagebreak[3]\begin{equation}}
\newcommand{\ee}{\end{equation}}
\newcommand{\ba}{\nopagebreak[3]\begin{eqnarray}}
\newcommand{\ea}{\end{eqnarray}}
\DeclareFontFamily{U}{rsfs}{}         
\DeclareFontShape{U}{rsfs}{m}{n}{<5> rsfs5 <6><7> rsfs7          
  <8><9><10><10.95><12><14.4><17.28><20.74><24.88> rsfs10}{}     
\DeclareMathAlphabet{\mathfs}{U}{rsfs}{m}{n}                     
                               
\def\pb#1{\rlap{\lower1.5ex\hbox{$\longleftarrow$}}{#1}}
\def\dpb#1{\rlap{\lower1.5ex\hbox{$\Longleftarrow$}}{#1}}
\def\spb#1{\rlap{\lower1.5ex\hbox{$\leftarrow$}}{#1}}
\def\sdpb#1{\rlap{\lower1.5ex\hbox{$\Leftarrow$}}{#1}}

\definecolor{blue}{rgb}{0,0,1}
\definecolor{green}{rgb}{0,1,0}
\definecolor{red}{rgb}{1,0,0}
\definecolor{vio}{rgb}{1,0,1}
\definecolor{ama}{rgb}{1,1,0}

\makeatletter
\@addtoreset{equation}{subsection}
\@addtoreset{section}{part}
\makeatother
\def\be{\begin{equation}}
\def\ee{\end{equation}}
\def\ba{\begin{eqnarray}}
\def\ea{\end{eqnarray}}

\begin{document}

\title{The Gibbs paradox, Black hole entropy and the thermodynamics of isolated horizons}

\date{\today}

\author{Andreas G.A. Pithis}

\thanks{pithis@cpt.univ-mrs.fr, andreas.pithis@campus.lmu.de}

\affiliation{Centre de Physique Th\'eorique\footnote{Unit\'e
Mixte de Recherche (UMR 6207) du CNRS et des Universit\'es
AMI, AMII, et du Sud Toulon-Var}, Campus de Luminy, 13288
Marseille, France\\ and\\ Arnold-Sommerfeld-Center for Theoretical Physics, Ludwig-Maximilians-Universit\"at, Theresienstr. 37, 80333 M\"unchen, Germany}

\begin{abstract}
This letter presents a new, solely thermodynamical argument for considering the states of the quantum isolated horizon of a black hole as distinguishable. We claim that only if the states are distinguishable, the thermodynamic entropy is an extensive quantity and can be well-defined. To show this, we make a comparison with a classical ideal gas system whose statistical description makes only sense if an additional $1/N!$-factor is included in the state counting in order to cure the Gibbs paradox. The case of the statistical description of a quantum isolated horizon is elaborated, to make the claim evident.
\end{abstract}


\maketitle

\subsection*{}
As solutions of the gravitational field equations teach us, a star having a sufficiently large mass collapses beyond its Schwarzschild radius and will continue to collapse completely into a gravitational singularity at the end of its lifetime. This renders the spacetime geodesically incomplete. The star disappears and leaves a black hole behind, whose size is given by the radius of the event horizon, i.e. the Schwarzschild radius, which covers the singularity. The event horizon is the frontier of all events, which in principle could be observed by an external observer - all inner events are concealed from the latter. The existence of these tantalizing objects is supported by an ever-increasing number of astrophysical observations. However, the motivation to attend to a more fundamental account of the gravitational field is driven firstly by the fact, that these singularities are unphysical divergences of the gravitational field and hence the predictability of classical general relativity for the description of the interior structure of the black hole breaks down as implied by the singularity theorems \cite{ST}. Secondly, dimensional arguments suggest, that one can't neglect quantum effects near these singularities.

Furthermore, based on its gravitational properties, it is possible to assign thermodynamic quantities to black holes. Concretely, this is accomplished through relating the horizon area to entropy, the black hole surface gravity to temperature and the black hole mass to energy, whose nomological relation is expressed by means of the four laws of black hole mechanics \cite{BCH}. Hence, these laws suggest a close analogy to the laws of thermodynamics. Additionally, calculations by means of semi-classical quantum field theory on curved spacetime predicting that quantum mechanically a black hole radiates like a black body with a temperature proportional to the surface gravity of the black hole, clarified the right proportionalities between the above related quantities. This gives rise to the Bekenstein-Hawking formula for the relation between entropy and area, 
\be\label{BeHawF}
S=k_B~(c^3/\hbar G)~A/4=k_B~A/(4\ell_p^2)
\ee
expressing a remarkable and intriguing relation between the geometry of spacetime, gravity, quantum theory and thermodynamics \cite{Hawking, beke}. Taking the microscopic underpinning of thermodynamics
through statistical mechanics seriously, the question arises, how this entropy can be microscopically explained. It is expected that it arises from the microstates of an underlying fundamental theory of quantum gravity describing the quantum structure of the horizon geometry.

It has been established for a while, that it is possible to describe black holes in equilibrium locally by the so-named isolated horizon boundary condition \cite{IsolatedHorizonFramework}. Physically this amounts to having no fluxes of matter and/or gravitational energy across this horizon. The isolated horizon framework is motivated by the fact, that the definition of a black hole as a spacetime region of no escape is a global one, i.e., it requires knowledge of the entire spacetime and further that it is in equilibrium. Consequently, it does not appear to be useful for describing local physics. This problem is solved by introducing the quasi-local notion of an isolated horizon, which is compatible with the laws of black hole mechanics. The quantum geometric handling of spacetimes with such an isolated horizon by means of Loop Quantum Gravity (LQG) techniques leads to the description of the black hole quantum degrees of freedom (d.o.f.) described by a Chern-Simons (CS) theory on a punctured $2$-sphere \cite{ABK,SU2,SU2rev}. The quantum geometry of the bulk is given by a spin network, whose graph pierces the horizon surface yielding the punctures, representing the quantum excitations of the gravitational field of the horizon. The dimension of the Hilbert space of this CS-theory can be used to compute its entropy, which interestingly agrees with the Bekenstein-Hawking entropy-area relation. In order to conduct a statistical mechanical analysis of such a system, one needs a proper notion of quasi-local energy of the horizon, whose introduction was succeeded in \cite{APE,espera}. With this at hand, the statistical mechanics of quantum isolated horizons was studied in the microcanonical, canonical and grandcanonical ensembles done in \cite{espera, Pranzetti}. In the present letter this recent study is used to analyze the impact of the distinguishability of the quantum d.o.f. on the extensivity of the entropy function.

To this aim, the letter is organized as follows. The first section summarizes the basic statistical mechanics of a classical ideal gas, focusing on the differences, which arise when particles are distinguishable or indistinguishable. This leads us to the famous Gibbs paradox and we will revise the notion of the Maxwell-Boltzmann statistics. The second section initially gives a summary of the state counting of the quantum d.o.f. of an isolated horizon, switching then to the discussion of the properties of the entropy. The third section closes the letter with a discussion and concluding remarks concerning the distinguishability of the quantum horizon states. 

\subsection*{I. Classical ideal Gas and the Gibbs paradox}
Let us briefly revise the derivation of the entropy $S$ of a classical ideal gas in the canonical ensemble and discuss the Gibbs paradox following \cite{StatMech}. The canonical partition function $Z(T,V,N)$ reads
\be
Z(T,V,N)=\sum_u exp\biggl(-\beta\sum_{i=1}^{3N}\frac{p_i^2}{2m}\biggr),
\ee
wherein a classical microstate of this gas is specified by $u=(x_1,...,x_{3N},p_1,...,p_{3N})$. It is possible to exchange the sum over the microstates $\sum_u$ by an integral over phase space $\frac{1}{N!h^{3N}}\int d\Gamma$, where $d\Gamma=d^{3N}x d^{3N}p$ and the well-known Gibbs factor $1/N!$ accounts for the indistinguishability of the particles. This is related to the fact, that the particles are non-localized. If they were localizable, one could distinguish them by their sites.\newline
The calculation gives
\be 
Z(T,V,N)=\frac{Z_1(T,V,1)^N}{N!}, Z_1\equiv V\biggl(\frac{2\pi mk_BT}{h^2}\biggr)^{3/2},
\ee 
where $Z_1$ denotes the one-particle-partition function. Using the expression for the free energy $F=-k_BT\mathrm{\log}Z$, the entropy $S(T,V,N)=-(\frac{\partial F}{\partial T})_{V,N}$ and Stirling's formula one arrives at
\be
S(T,V,N)=Nk_B(\mathrm{\log}V-\mathrm{\log}N+\frac{3}{2}\mathrm{\log} T+\alpha),
\ee
with $\alpha=3/2 \mathrm{\log}(2\pi mk_B/h^2)+5/2$.\newline
Now consider a canister of volume $V$ containing a classical ideal gas of $N$ identical particles with entropy $S$. Let us divide the volume through insertion of a wall into two parts $V_1$ and $V_2$. This is a special case of the mixing entropy problem. The particle density in each volume is certainly constant
\be
\rho=N/V=N_1/V_1=N_2/V_2=const..
\ee
This gives 
\be
S_i=N_ik_B\biggl(-\mathrm{\log}\rho + 3/2 \mathrm{\log} T + \alpha\biggr).
\ee
Adding up the entropy contributions of each part, one obtains $S_1+S_2=S_{tot}\stackrel{!}{=}S$.\newline
Hence,
\be\label{DeltaSri}
\Delta S=S-S_{tot}=0,
\ee
since the insertion of the wall is a reversible process.\newline
Now let us compare this elementary result with the case, where the particles are considered to be distinguishable. Then the Gibbs factor $1/N!$ is left out and one computes the entropy to give
\be
S_i=N_ik_B (\mathrm{\log} V_i + 3/2 \mathrm{\log} T + \alpha-1).
\ee
Using this result, one finds that 
\be
S_{tot}=S_1+S_2\neq S
\ee 
and can explicitly be brought in the following form
\be
\Delta S\approx k_B\mathrm{\log} \biggl(\frac{(N_1+N_2)!}{N_1!N_2!}\biggr)\neq 0,
\ee
which is quite different from the above result (\ref{DeltaSri}). Here the entropy is simply not an extensive function, if the $1/N!$-factor is left out. Historically, Gibbs worked at first under the then common hypothesis, that the identical particles should be distinguishable by their trajectories in phase space and arrived at this paradoxical result. Including the factor made the entropy extensive and cured the problem. Gibbs' introduction was certainly ad hoc but was later properly justified by means of quantum mechanics.

The question arises, whether and how this argumentation can be applied to the case of quantum isolated horizons, considered to consist of a gas of topological defects, also called punctures. These are usually taken to be distinguishable and the Maxwell-Boltzmann statistics is applied, which can tempt one to think that the Gibbs paradox could also show up in this model.

Before rushing ahead, let us generalize the above discussion following \cite{StatMech}, particularly with regard to the work done in \cite{espera}. Let us consider a classical gaseous system of $N$ non-interacting particles in a canister of volume $V$. The total energy, which is shared among the particles is $E$. The underlying statistical distribution is the Maxwell-Boltzmann statistics. If $V$ is large, the energy spectrum is divided into energy cells. The $j$th cell has an average energy $\epsilon_j$ and is occupied by $n_j$ particles. For generality let $d_j$ denote the number of levels in the $jth$ cell, also called the degeneracy of the level $j$. The total number of particles and the total energy constrain the distribution set $\{n_j\}$, namely
\be
N=\sum\limits_{j} n_j,~~~~~~~~E=\sum\limits_j n_j\epsilon_j.
\ee
Let $W^{MB}_{\{n_{j}\}}$ denote the number of distinct microstates belonging to the distribution set $\{n_j\}$ and let $w(j)$ denote the number of distinct microstates belonging to the $j$th cell, then
\be\label{WMB}
W^{MB}_{\{n_{j}\}}=\prod_jw(j)=N!\prod_j\frac{(d_j)^{n_j}}{n_j!}.
\ee
Any of the $n_j$ objects are put independently of each other into any of the levels $d_j$. The resulting states are considered to be distinct and the number of these states is given by $d_j^{n_j}$. Additionally, an exchange of particles in the same $\epsilon_j$-cell shall not give new microstates and therefore one has the product of the $n_j!$-factors. The $N!$-factor shall express, that an exchange of two particles in different $\epsilon_j$-cells gives new, physically distinct microstates. This reflects the fact, that classical particles are mutually distinguishable, though being identical. On the other hand, if one assumes that an exchange of two particles in different $\epsilon_j$-cells does not give a new microstate, the $N!$-factor is left out. The particles are then considered to be indistinguishable and the number of distinct microstates belonging to the distribution set $\{n_j\}$ is given as
\be\label{Wc}
W^{cMB}_{\{n_{j}\}}=\prod_j\frac{(d_j)^{n_j}}{n_j!},
\ee
where "$cMB$" expresses the counting corrected by the Gibbs factor $1/N!$.
\newline
The entropy of the system is then given by
\be\label{allgEntropie}
S(E,N,V)=k_B~\mathrm{\log}\Omega(E,N,V)=k_B~\mathrm{\log}\biggl(\sum\limits_{\{n_j\}}^{-}W_{\{n_{j}\}}\biggr),
\ee
\newline
where $\Omega(E,N,V)$ is the number of distinct microstates, which are accessible to the system in the macrostate $(E,N,V)$. The "${-}$" on the summation symbol implies that the sum is restricted to distribution sets $\{n_j\}$, which obey the conditions on $N$ and $E$. It is possible to approximate the entropy to
\be
S(E,N,V)=k_B~\mathrm{\log}(W_{\{n_{j}^{*}\}}),
\ee
wherein the distribution set $\{n_{j}^{*}\}$ maximizes $W_{\{n_{j}\}}$ under the conditions that $E,N=const.$, which are enforced by two Lagrange multipliers. The $n_j^{*}$'s are interpreted as the most probable values of the distribution numbers $n_j$ and are given for distinguishable particles by
\be
n_j^{*}=N d_j e^{-\beta(\epsilon_j-\mu)}
\ee
and for indistinguishable particles by
\be
n_j^{*}= d_j e^{-\beta(\epsilon_j-\mu)},
\ee
respectively. Therein $\beta=1/k_BT$ and $\mu$ is the chemical potential. In the following we will set Boltzmann's constant $k_B$ to $1$.

\subsection*{II. Entropy of quantum isolated horizons and the distinguishability of punctures}
In the second part of this letter, the statistical mechanical properties of quantum isolated horizons will be treated following closely \cite{espera}. Necessary results are summarized to give a pedagogical access to the analysis of the extensivity property of the entropy function of the discussed model.

In Loop Quantum Gravity one effectively considers the quantum gravitational d.o.f. of the quantum isolated horizon to consist of a gas of topological defects. These are punctures, which arise from the horizon being pierced by the edges of the spin network, which describes the bulk quantum geometry. The punctures are described by means of Chern-Simons theories with $U(1)$- or $SU(2)$-gauge groups coupled to point-like sources, where the gauge connection has non-vanishing support. For the short discussion here, only the $SU(2)$ case will be considered. The microstate counting is then accomplished using for example the relation between the Hilbert space of a $SU(2)_k$ CS-theory\footnote{$k$ denotes the level of the CS-theory. Usually, but not necessarily, one considers the level to scale with the macroscopical classical horizon area $k\propto a_H/\ell_p^2$ setting $k\to\infty$. The case where $k$ is fixed can be found in \cite{SU2rev,PP, Mitra, PranzettiPolo}.} on the bounding isolated horizon with the space of conformal blocks of the WZWN-model living on the bounding $2$-sphere $S^2$ \cite{KaulMajumdar}. It can also be done directly in the $SU(2)_k$ CS-theory, without having to use the techniques of CFT \cite{SU2Character,ABK,BE,PP,Me,GM,SU2} and additionally, was achieved using the representation theory of the quantum deformed $SU_{q}(2)$, where $q$ is a non-trivial root of unity \cite{SU2rev}. All these approaches arrive at the following master formula. It gives the number of microstates $W_{\{n_{j}\}}$ associated to a quantum configuration $\{n_j\}$, given by the number of punctures $n_j$ labelled by an irreducible representation $\rho_j$
\be
W_{\{n_{j}\}}=\frac{N!}{\prod_j(n_j!)}~\frac{2}{k+2} \sum_{l=0}^{k/2}\mathrm{sin}^2\biggl(\frac{(2l+1)\pi}{k+2}\biggr)\prod_{j} d^{n_j}_j(l),
\ee
wherein
\be
d_j(l)\equiv\biggl[\frac{\mathrm{sin}(\frac{(2j+1)(2l+1)\pi}{k+2})}{\mathrm{sin}(\frac{(2l+1)\pi}{k+2})}\biggr].
\ee
By taking the naive limit, where $k\to \infty$ and neglecting the next-to-leading order term\footnote{This term gives rise to the $-3/2\mathrm{\log}A$ correction of the entropy (e.g., calculated in \cite{KaulMajumdar, Mitra, logcorrection}).}, one obtains up to a numerical factor
\be\label{W}
W_{\{n_{j}\}}=N!\prod_j\frac{(2j+1)^{n_j}}{n_j!},
\ee
which takes the form of a typical Maxwell-Boltzmann distribution (\ref{WMB}). It gives the number of distinct microstates associated with the given configuration $\{n_j\}$, where one has $(2j+1)$ different projections of $j$, which gives rise to $\prod_j(2j+1)^{n_j}$. The factor $N!/{\prod_j (n_j!)}$ is explained as in the first section.\newline
The thermodynamical properties of this system can only be analyzed, if one introduces a notion of quasi-local energy of the horizon. Isolated horizons don't allow for a unique notion of quasi-local energy, but with the help of a local first law for isolated horizons \cite{APE} and an additional physical input, e.g., the Schwarzschild spacetime, a unique notion of quasi-local energy can be given \cite{APE,espera}. The natural quasi-local energy associated with a stationary observer in Schwarzschild spacetime is computed there with the help of the Komar-mass integral giving
\be\label{E}
E=\frac{A}{8\pi \ell},
\ee
wherein $\ell$ denotes the proper distance from the horizon, which is inversely proportional to the local surface gravity $\bar{\kappa}$. Beyond that, a proper notion of temperature for the quantum isolated horizon was given as
\be\label{loctemp}
T_U=\frac{\ell_p^2}{2\pi \ell},
\ee
where $\ell_p$ is the Planck length. It is the temperature, which is measured by a family of local non-rotating stationary observers, that hover outside the horizon at proper distance $\ell$ to it \cite{espera,APE,Pranzetti}.\newline
By using the operator version of (\ref{E}) with the definition of the area spectrum of LQG \cite{LQG}, one obtains the scaled isolated horizon area spectrum as the energy spectrum of the punctures
\be
\widehat H|\{n_j\}\rangle=\biggl(\frac{\ell^2_g}{\ell} \sum_{j}n_j \sqrt{j (j+1)}\biggl)|\{n_j\}\rangle,
\ee
wherein $\ell_g^2=\gamma\ell_p^2$ and $\gamma$ is the Immirzi-parameter.\newline
With the energy (\ref{E}) and the total number of punctures $N$, the statistical treatment within the microcanonical ensemble is at hand using the number of distinct microstates $W_{\{n_j\}}$ for a quantum configuration $\{n_j\}$ as in (\ref{W}). With this, one searches for the dominant configuration $n_j^{*}$, which maximizes the entropy, while considering the two constraints
\be
N=\sum_j n_j,~~~~~~~~\sum_j n_j\sqrt{j(j+1)}=\frac{A}{8\pi \ell_g^2},
\ee
enforced by the Lagrange multipliers $\sigma$ and $\lambda$, respectively.\newline
This yields the dominant configuration
\be
n_j^{*}=N (2j+1) e^{-\lambda\sqrt{j(j+1)}-\sigma}
\ee
with the consistency condition
\be
1=e^{-\sigma}\sum_j(2j+1)e^{-\lambda\sqrt{j(j+1)}},
\ee
from which $\sigma(\beta,\gamma)$ can be computed.\newline
Using $\beta=(\partial S/\partial E)_N$, one finds $\lambda=\beta l_g^2/l$. This is used to give the expression for the entropy
\be
S=\mathrm{\log}(W(\{n_{j}^{*}\}))=\lambda\frac{A}{8\pi \ell_g^2}+\sigma N.
\ee
If one plugs the local temperature (\ref{loctemp}) into this expression, one obtains
\be\label{BHent} 
S=A/(4\ell_p^2)+\sigma(\gamma)N,
\ee
being compatible with the semi-classical Bekenstein-Hawking area law (\ref{BeHawF}), as the second summand only expresses the quantum hair correction due to the quantum geometry of the isolated horizon \cite{espera}.\newline
Now this system shall be discussed in the canonical ensemble, in which the prime objective of this letter is most easily seen. For this, we need the canonical partition function 
\be
Z=\sum^{-}_{\{n_{j}\}}N!\prod_j\frac{(2j+1)^{n_j}}{n_j!}e^{-\beta n_j \epsilon_j}
\ee
with $\epsilon_j=\ell_g^2/\ell~\sqrt{j(j+1)}$. By means of the multinomial theorem the partition function for distinguishable punctures is rewritten and gives
\be
Z(N,T)\equiv(Z_1)^N=\biggl(\sum_j(2j+1)e^{-\beta \epsilon_j}\biggr)^N,
\ee
where $Z_1$ denotes the one-particle-partition function.\newline
Using the standard formula for the average energy
\be\label{averageE}
\langle E \rangle=-\partial_{\beta}\mathrm{\log}Z=-N\frac{Z_1^{'}}{Z_1},
\ee
and with 
\be\label{Entropymaster}
S=-\beta^2\partial_{\beta}\biggl(\frac{1}{\beta}\mathrm{\log}Z\biggr)=\beta\langle E\rangle+\mathrm{\log}Z
\ee
one arrives at
\be\label{Sdist}
S=N\biggr(\mathrm{\log}Z_1-\beta\frac{Z_1^{'}}{Z_1}\biggr).\footnote{If one computes $S(T_U)$, one is directly lead to $S=A/(4\ell_p^2)+\sigma(\gamma)N$, being in agreement with the previously recovered result in the microcanonical ensemble (\ref{BHent}).}
\ee
This is certainly an extensive function in $N$, which is consistent with the requirement that the entropy of a system is equal to the sum of the entropies of its parts. The corresponding picture to the partition of a canister with volume $V$ would be the admittedly exotic case of merging two black holes with common initial temperature $T$, each described by its quantum isolated horizon and respective quantum statistical description. According to (\ref{Sdist})
\be\label{totalentropydist} 
S_{tot}=S_1+S_2=(N_1+N_2)\biggl(\mathrm{\log}Z_1-\beta\frac{Z_1^{'}}{Z_1}\biggr)
\ee
and then
\be
\Delta S=S-S_{tot}=0,
\ee
which expresses, that \emph{theoretically} this should be a reversible process, when neglecting all sorts of radiative processes.
Now let us consider the other case, when the punctures are taken to be indistinguishable.
The canonical partition function simply has to get modified by the Gibbs factor, yielding
\be
Z(T,N)=\frac{(Z_1)^N}{N!}.
\ee
Crucially, the application of the average energy formula (\ref{averageE}) gives also
\be
\langle E\rangle=-N\frac{Z_1^{'}}{Z_1}.
\ee
and together with (\ref{Entropymaster}) and Stirling's formula, one obtains
\be
S=N\biggl(\mathrm{\log}Z_1-\beta\frac{Z_1^{'}}{Z_1}-\mathrm{\log}N\biggr),
\ee
which is certainly not an extensive function in $N$. Consequently, one obtains exactly
\be
S_{tot}=S_1+S_2=eq.(\ref{totalentropydist})-\mathrm{\log}(N_1!N_2!)
\ee
and with this the entropy difference yields
\be
\Delta S=S-S_{tot}=\mathrm{\log}\biggl(\frac{N_1!N_2!}{N!}\biggr)\leq0,
\ee
violating the second law of thermodynamics.

\subsection*{III. Discussion and conclusion}

We have used the statistical mechanical framework of quantum isolated horizons, which was introduced in \cite{APE, espera, Pranzetti}, to study the property of the extensivity of the entropy function. It was explicitly shown, that when the quantum statistics of the isolated horizon is accounted for by the Maxwell-Boltzmann statistics for distinguishable entities (\ref{WMB}, \ref{W}), a well-defined thermodynamic entropy is obtained. In this way, our analysis provides further support to the paradigm - and this time from a thermodynamical perspective - that the quantum states of the isolated horizon are distinguishable from one another. 

In \cite{Kiefer} the question was raised, whether the Gibbs paradox would arise in the context of the LQG treatment of black hole entropy due to the fact that the fundamental entities building up the horizon are distinguishable. However, the analysis of the extensivity of the entropy in this letter shows, that if the punctures were indistinguishable and their statistical distribution was captured by the Maxwell-Boltzmann statistics modified by the additional Gibbs factor (\ref{Wc}), one would not arrive at a properly defined notion of entropy. To guarantee the extensivity of the entropy function, the punctures have to be distinguishable. This is much like a model approximating a solid, consisting of a system of independent and localized particles, where in the thermodynamic limit $S\propto N$. Its statistical properties are of course different from the one of a classical ideal gas, where the Gibbs factor occurs and a very different one-particle-partition function due to the different microdynamics is used \cite{StatMech}.

Additionally, in \cite{Krasnov, APS} it was argued that the consideration of a counting based on indistinguishable punctures, would give rise to an entropy-area relation as $S\propto {A}^t$, with $t<1$, being in conflict with the laws of black-hole mechanics and additionally the entropy would not be an extensive function of the area.

We see that the particular form $S\propto A$ and the extensivity of the entropy with respect to $A$ and $N$ can be understood as benchmarks for the form of the horizon quantum statistics and indeed they are met, when the quantum horizon states are distinguishable from one another. In this light, it seems helpful to revise the underlying conceptual arguments for the distinguishability below. 

In this work we recollected, that within the isolated horizon framework \cite{IsolatedHorizonFramework}, the horizon is treated as an inner boundary of space, which suffices the isolated horizon boundary condition. As a consequence, one obtains a surface term proportional to the action of Chern-Simons theory in the gravitational action. Then one quantizes the bulk and horizon d.o.f. separately and uses afterwards the quantum horizon boundary condition 
\be\label{IHBC} 
\epsilon^{ab}\hat{F}^i_{ab}+\frac{4\pi}{k}\sum_{p\in\mathcal{P}}\delta(x,x_p)\hat{J}^i(p)=0,
\ee
in order to couple them again \cite{ABK, SU2, PP, PranzettiPolo, Commentoncoupling}. The level of the CS-theory is denoted by $k=a_H/(2\pi\gamma(1-\gamma^2)\ell_p^2)$, $\mathcal{P}=\{(p_1,j_{p_1}),...,(p_N,j_{p_N})\}$ runs over all finite puncture sets and $\hat{F}^i_{ab}$ is the quantization of the curvature $2$-form of the Ashtekar-connection $A^i_a$ being pulled-back to the horizon surface $S$ and is therefore the curvature of the $SU(2)$ CS-connection. $i=1,2,3$ indicates, that the internal index $i$ carries a representation of the Lie algebra $su(2)$, whereas $a,b=2,3$ refers to the pulled-back spatial indices, which are the $(\theta,\phi)$- coordinates on the $2$-sphere. At each puncture $p$ the well-known angular momentum algebra $[\hat{J}^i(p),\hat{J}^j(p)]=\epsilon^{ij}_k\hat{J}^k(p)$ holds, where the operators $\hat{J}^i(p)$ act as point sources for the $SU(2)$ CS-theory. It should be emphasized that both sides of the equation are $su(2)$-valued, already telling us that the bulk and surface d.o.f. cannot be arbitrarily coupled \cite{ABK, SU2, Commentoncoupling}. The states, which satisfy (\ref{IHBC}) are elements of 
\be
\mathcal{H}_{phys}=\biggl(\bigoplus_\mathcal{P}\mathcal{H}^{\mathcal{P}}_{B}\otimes\mathcal{H}^{\mathcal{P}}_{S}\biggr)/G,
\ee
where $\mathcal{H}^{\mathcal{P}}_{B}$ denotes bulk space of states. One denotes by $G$ internal $SU(2)$-transformations, diffeomorphisms, which preserve the surface and eventually motions, generated by the Hamiltonian constraint $H$. The lapse is restricted to vanish on the horizon and thus $H$ is only imposed in the bulk. Finally, $\mathcal{H}^{\mathcal{P}}_{S}$ is the Hilbert space of the above introduced Chern-Simons theory on the punctured sphere, which is a topological quantum field theory (TQFT) \cite{TQFT}.

The early computations of black hole entropy in the framework of LQG were much shaped by paradigms, which were set in \cite{Rovelli}. It is instructive to recapitulate a group theoretical argument given there, that one is restricted to consider only horizon diffeomorphisms, that do not arbitrarily mix the intersections (punctures) between the bulk spin network and the horizon surface. To this aim, let us now solely concentrate on the diffeomorphisms of the horizon surface, neglecting the other gauge transformations. Taking the quotient by these diffeomorphisms leads to the identification of any two Hilbert spaces, belonging to the puncture sets $\mathcal{P}$, which can be mapped into one another through a diffeomorphism on the surface. No reference to the position of the punctures on the horizon has to be made as a consequence of imposing the diffeomorphism constraint. To understand the distinguishability, one has to investigate to what the space $(\mathcal{H}^{\mathcal{P}}_{B}\otimes \mathcal{H}^{\mathcal{P}}_{S})/G$ is isomorphic. It is possible to differentiate two cases, depending on whether the action of $G$ is free or not. Since their explanation is similar, we briefly dwell on the latter case. If $G$ does not act freely on $\mathcal{H}^{\mathcal{P}}_{B}\otimes \mathcal{H}^{\mathcal{P}}_{S}$, one has 
\be
(\mathcal{H}^{\mathcal{P}}_{B}\otimes \mathcal{H}^{\mathcal{P}}_{S})/G\cong \mathcal{H}^{\mathcal{P}}_{B}/G \otimes \bigcup_{a\in \mathcal{H}^{\mathcal{P}}_{B}/G}\mathcal{H}^{\mathcal{P}}_{S}/G_a,
\ee
where $G_a$ is the stabilizer of $a$ with respect to $G$. Let $G$ act on $(a,b)\in\mathcal{H}^{\mathcal{P}}_{B}\otimes \mathcal{H}^{\mathcal{P}}_{S}$ like $g(a,b)=(a',b')$. If $(a,b)$ and $(a',b')$ are in the same $G$-orbit, then this can only be achieved, if one requires that the surface state $b'\in \mathcal{H}^{\mathcal{P}}_{S}/G_a$. Any $(a'',b'')$ with $b''\notin \mathcal{H}^{\mathcal{P}}_{S}/G_a$ does not lie in the same $G$-orbit as $(a,b)$ and is distinct from the latter. Consequently, equivalent states, i.e. states in the same $G$-orbit, can be discriminated from states in different orbits. Nevertheless, all these orbits are compatible with the same total area of the horizon. Physically this means, that diffeomorphic horizon configurations cannot be distinguished from an external point of observation, e.g., from the perspective of the above introduced family of local stationary obervers. Hence, physical states are defined as equivalence classes under these diffeomorphisms and from the statistical point of view one actually has to count such different equivalence classes, which correspond to different microstates and that are accessible to the system in the macrostate $(E,N)$. Each different microstate has a microscopically distinct effect on the exterior (bulk) of the black hole, where the local observer resides and is able to differentiate between them. This implies, that any operation changing the order of the configuration $\{n_j\}$, gives a distinct microstate. Applied to the above argumentation, we saw that any permutation of punctures in the same representation $\rho_j$ does not give a new microstate, hence the $1/\prod_j(n_j!)$-factor. These states lie in the same orbit. On the other hand, any permutation exchanging punctures in different representations $\rho_{j_{1}},~\rho_{j_{2}}$ gives a distinct microstate, accounted for by the $N!$-factor. Together with the $(2j+1)$-degeneracy of each representation $\rho_j$ this gives the total number of distinct microstates (\ref{W}) associated to the quantum configuration of punctures $\{n_j\}$.

This view is also supported from another perspective, since in the recent rigorous attempt at providing a full definition of a quantum horizon from within LQG in \cite{Sahlmann}, the invariance of the quantum horizon state $\psi$ (being a solution to (\ref{IHBC})) under diffeomorphisms, which leave the punctures fixed, was explicitly shown. It was noted in \cite{Sahlmann, Pranzetti}, that this symmetry gets broken, if one interchanges two differently labelled punctures, while leaving the other ones invariant. This goes nicely in hand with the statement, that the presence of a boundary in a theory can break the gauge invariance, turning gauge d.o.f. into new physical boundary d.o.f., described by an induced TQFT on the boundary and consequently, that these new "would-be gauge" d.o.f. can lead to a drastic alteration of its corresponding Hilbert space \cite{Carlip}.

Finally, it is also helpful to pay carefully attention to the distinction between the notions of identicality and indistinguishability in quantum physics \cite{StatMech}. Quantum entities are denoted as identical, if they have their state-independent or intrinsic properties in common, which are certain exactly equal observable parameters. These are for example constant properties such as charges, spin and rest mass. Whether a property is intrinsic or not, is a system-relative issue. Non-identical objects are always distinguishable by their different intrinsic or state-independent properties. Now, identical entities are denoted as indistinguishable, if they obey the indistinguishability postulate, which states that \emph{all} observables $\hat{O}$ must commute with \emph{all} permutations $P$ of the entities, i.e. $[\hat{O},P]=0$. The corresponding microstates remain unchanged under a permutation of two objects. In contrast, if a permutation that interchanges entities in two different single-entity states leads to a physically distinct microstate of the system, then each of these microstates has to be taken care of in the state counting procedure separately. In the case of the horizon punctures, one deals with identical and nonetheless distinguishable entities. The puncture states are labelled with different $SU(2)$-representations and the label $j$ is clearly a state-dependent property. Though $[\hat{A},P]=0$, where $P$ denotes a puncture permutation, interchanging punctures with different $j$'s has nonetheless a microscopically distinct effect on the bulk, i.e. $[\hat{J}^i(p),P]\neq 0$. It ensues, that a microstate is changed by such a puncture permutation into another one and the former can be discriminated from the latter. Consequently, a statistical distribution has to reflect this dinstinguishability.

{\bf Acknowledgements.} I am deeply grateful to C. Rovelli for having me in Marseille, for letting me into the very stimulating ambience of the group and thus giving me the chance to explore aspects of quantum gravity. I also want to express honest thanks to A. Perez for being so helpful and patient with my questions and to S. Hofmann for making it possible for me to be abroad for this work and for his support in the mean time. I am also grateful to M. Han, D. Pranzetti, W. Wieland, H.-Ch. Ruiz, P. Tomov, M. Haeberlein, J. Th\"urigen, F. G\"ade and J. Hofereiter for helpful remarks on an earlier version of this work.

\end{document}